# Title: Chemical Partitioning at Crystalline Defects in PtAu as a Pathway to Stabilize Electrocatalysts

**Authors:** Xuyang Zhou[1,2,*], Olga Kasian[1,3,4,*], Ting Luo[1], Se-Ho Kim[1], Chenyu Zhang[5], Siyuan Zhang[2], Subin Lee[2], Gregory B. Thompson[6], Gerhard Dehm[2], Baptiste Gault[1,7,*], Dierk Raabe[1,*]

**Affiliations:**

[1] Department of Microstructure Physics & Alloy Design, Max-Planck-Institut für Eisenforschung GmbH, 40237 Düsseldorf, Germany.

[2] Department of Structure & Nano- / Micromechanics of Materials Max-Planck-Institut für Eisenforschung GmbH, 40237 Düsseldorf, Germany.

[3] Helmholtz-Zentrum Berlin GmbH, Helmholtz Institut Erlangen-Nürnberg, Cauerstr. 1, 91058 Erlangen, Germany

[4] Department of Materials Science and Engineering, Friedrich-Alexander-Universität Erlangen-Nürnberg, 91058 Erlangen, Germany

[5] Applied and Engineering Physics, Cornell University, 14853 Ithaca, USA

[6] Department of Metallurgical Materials Engineering, The University of Alabama, 35487 Tuscaloosa, USA

[7] Department of Materials, Royal School of Mines, Imperial College London, London, UK

[*]Corresponding to: x.zhou@mpie.de; olga.kasian@helmholtz-berlin.de; b.gault@mpie.de; d.raabe@mpie.de;

**Abstract:** Dissolution of electrocatalysts during long-term and dynamic operation is a challenging problem in energy conversion and storage devices such as fuel cells and electrolyzers [1,2]. To develop stable electrocatalysts, we adopt the design concept of segregation engineering [3,4], which uses solute segregation prone to electrochemical dissolution at internal defects, *i.e*., grain boundaries and dislocations. We showcase the feasibility of this approach by stabilizing a model Pt catalyst [5] with an addition of more noble Au [6] (approximately 5 at.%). We characterized the defects' nanoscale structure and chemistry, and monitored the electrochemical dissolution of Pt and PtAu alloys by on-line inductively coupled plasma mass spectrometry [7-10]. Once segregated to defects, Au atoms can stabilize and hence passivate the most vulnerable sites against electrochemical dissolution and improve the stability and longevity of the Pt electrocatalysts by more than an order of magnitude. This opens pathways to use solute segregation to defects for the development of more stable nanoscale electrocatalysts, a concept applicable for a wide range of catalytic systems.

**Main Text:** The demand for clean energy massively exceeds its supply on the world market. Chemical energy buffering plays a central role in that context. In particular, environmentally friendly electrochemical technologies, such as fuel cells or water electrolyzers, powered by renewables can ensure transition of our energy systems to renewable sources in multiple sectors. The current challenge in developing electrochemical devices for energy conversion and storage is to reduce their fabrication and operational costs, including by ensuring long-term stable operation of the catalysts [1,2]. Despite major progress made in the development of low-cost, noble metals-free electrocatalysts [2,11,12], they yield lower durability compared to the state-of-the-art catalysts made of Pt-group metals [13,14]. Superior resistance towards corrosion makes Pt the most commonly used material in electrochemical applications. In terms of the balance between electrocatalytic activity and stability towards dissolution there is no other metal that could compete with Pt in electrocatalysis of the oxygen reduction reaction (ORR) and hydrogen evolution reaction (HER) in acidic media [5]. Still, even Pt is not completely immune to the harsh, dynamic operational conditions, typical for fuel cells and electrolyzers [10], and slowly degrades. Considering the high price and low abundance of Pt, understanding the underlying principles of degradation at atomic level and development of strategies to stabilize catalyst is therefore of high importance [15].

Designing the microstructure and composition of Pt-based catalysts improves their efficiency and stability, for example, increasing the reaction interfaces by porous nanostructures enhances the catalytic activity of Pt [16] and high ORR rates can be achieved by alloying, *e.g.*, by using PtNi, PtCo, PtCr alloys [17-21]. However, stability investigations on Pt alloys reported preferable leaching of the less noble component from the matrix of the catalyst [14,22]. On the one hand selective leaching processes lead to improvement of the catalytic activity because of formation of structures with enhanced surface area. On the other hand, such structures suffer from low mechanical and electrochemical stability. Substantial progress in understanding of Pt stability came from studies using electrochemical scanning flow cell connected to inductively coupled plasma mass spectrometry [7-9] and electrochemical characterization combined with transmission electron microscopy (TEM) [22-24]. The in-situ methods enabled quantification of Pt loss during operation and resulted in several dissolution mechanisms proposed in literature [13,14,25,26]. Most of the findings reveal that Pt is rather stable in conditions of constant anodic or cathodic polarization, but suffers from the dissolution under transient conditions, when the potential is changing over time. Such conditions naturally occur in the full cells and electrolyzers during the shutdown and startup cycles and play a crucial role in sustainable energy applications due to the intermittent nature of renewable energy sources.

Here, our goal is to improve the intrinsic stability of Pt electrocatalysts by using segregation engineering as an atomic-scale material design tool: solute decoration is exploited to manipulate the stability of defects such as dislocations, grain boundaries (GBs), and interfaces [3,4]. By tuning the chemical distribution of an alloying element at defects sites, we change the system's free energy and hence modify the dissolution process of electrocatalysts in operation [15]. Identifying a solute with a suited defect trapping energy at defects in the host metal, i.e. a dopant, qualifies this approach as a self-organized and defect-specific blending and protection method. We showcase our approach by mixing the more stable and more noble element Au [6] into Pt. Considering the scarcity of Au, we limited its content in the alloy to only 5 at.%. Our hypothesis is that segregation of Au at defects (a) protects Pt electrocatalysts from dissolution and (b) particularly passivates the undecorated defects where dissolution is expected to initiate. Insights into defect types and segregation chemistry are quantitatively resolved by advanced atomic-scale material characterization methods such as high-resolution scanning transmission electron microscopy (HR-STEM) and atom probe tomography (APT) [27-31]. We demonstrate that Pt-defects prone to dissolution are 'sealed' by segregation of the thermodynamically more stable Au [29,32], thereby improving durability, and propose a mechanistic understanding of the dissolution process. Our detailed study of the underlying structure-composition-property relationships open up pathways to use solute segregation to defects for the atomic scale design and synthesis of advanced, stable electrocatalysts for efficient and sustainable conversion and storage of energy.

**Results**

**Atomic-scale composition and structure**. A Pt-5(at.%)Au thin film alloy catalyst, referred to as PtAu in the following, was prepared by co-sputtering, followed by an annealing treatment at 773 K in a vacuum chamber for 4, 20 or 100 hours. Annealing triggers controllable diffusion, allowing systematic investigation from short-range solute diffusion and trapping at defects to clustering, capillary coarsening, and even phase separation [29,32].

We found that the as-deposited PtAu thin film appears to be compositionally homogenous, Fig. 1a-i. After an annealing treatment at 773 K for 100 hours, Au segregation is readily observed, Fig. 1a-ii. To characterize the nanoscale microstructure of the Pt and PtAu thin films, we employed four-dimensional

STEM (4D-STEM) to quantify the crystallographic texture, grain size, and GB types, see Extended Data Fig. 1. The Pt and PtAu thin films form a strong {111} texture. After an annealing treatment at 773 K for 100 hours, the grain size of the PtAu thin films increased from 75±41 to 118±58nm. We found a large fraction, approx. 30%, of low angle GB (LAGB) for all the studied thin films, see Extended Data Fig. 1. Fig. 1b shows the atomic structure and composition of a 6° - LAGB. HR-STEM and energy-dispersive X-ray spectroscopy (EDS) analyses show the direct correlation between the local defect structure and solute segregation. In Fig. 1b-i, the high-angle annular dark field (HAADF) image reveals two dislocations separated by approximately 2 nm. The Bragg filtered phase map [33] of the $(0\bar{2}2)$ planes provides a direct view of the insert planes near the dislocation cores, indicating that these are edge dislocations, Fig. 1b-ii. The calculated strain in the $(0\bar{2}2)$ direction, using the geometric phase analysis (GPA) method [33], is plotted in Fig. 1b-iii. This facilitates visualization of tensile and compressive zones near the dislocation cores [32], a feature which governs the chemical decoration. Au atoms are 4 % larger than Pt atoms, which drives Au segregation into the regions with positive hydrostatic stress at and near the dislocation cores. For the same region, we quantified the distribution of Au atoms by near-atomically-resolved EDS mapping, Fig. 1b-iv. This result matches the prediction of the Cottrell theory [34]. Interestingly, we found an asymmetric distribution of Au atoms around the dislocation core. We attribute this asymmetry to the elastic field overlap between adjacent dislocations.

A dependence of solute segregation on the GB structure is revealed by characterizing selected GBs in the 773K/100h PtAu thin film, see Fig. 1c. For the high angle GBs (HAGBs) shown in Fig. 1c-i, the solute segregation appears to be more continuous, without the periodic pattern observed in the LAGBs made of discrete dislocations. Nevertheless, local defects, such as facets, steps, and disconnections on the HAGBs, differ in the spectrum of segregation energies [35] and thus cause local differences in trapping, which finally lead to an inhomogeneous solute decoration along such interfaces [36-38], see the faceting of Σ3 (where Σ3 is the density of coincident sites) GBs in Fig. 1c-ii.

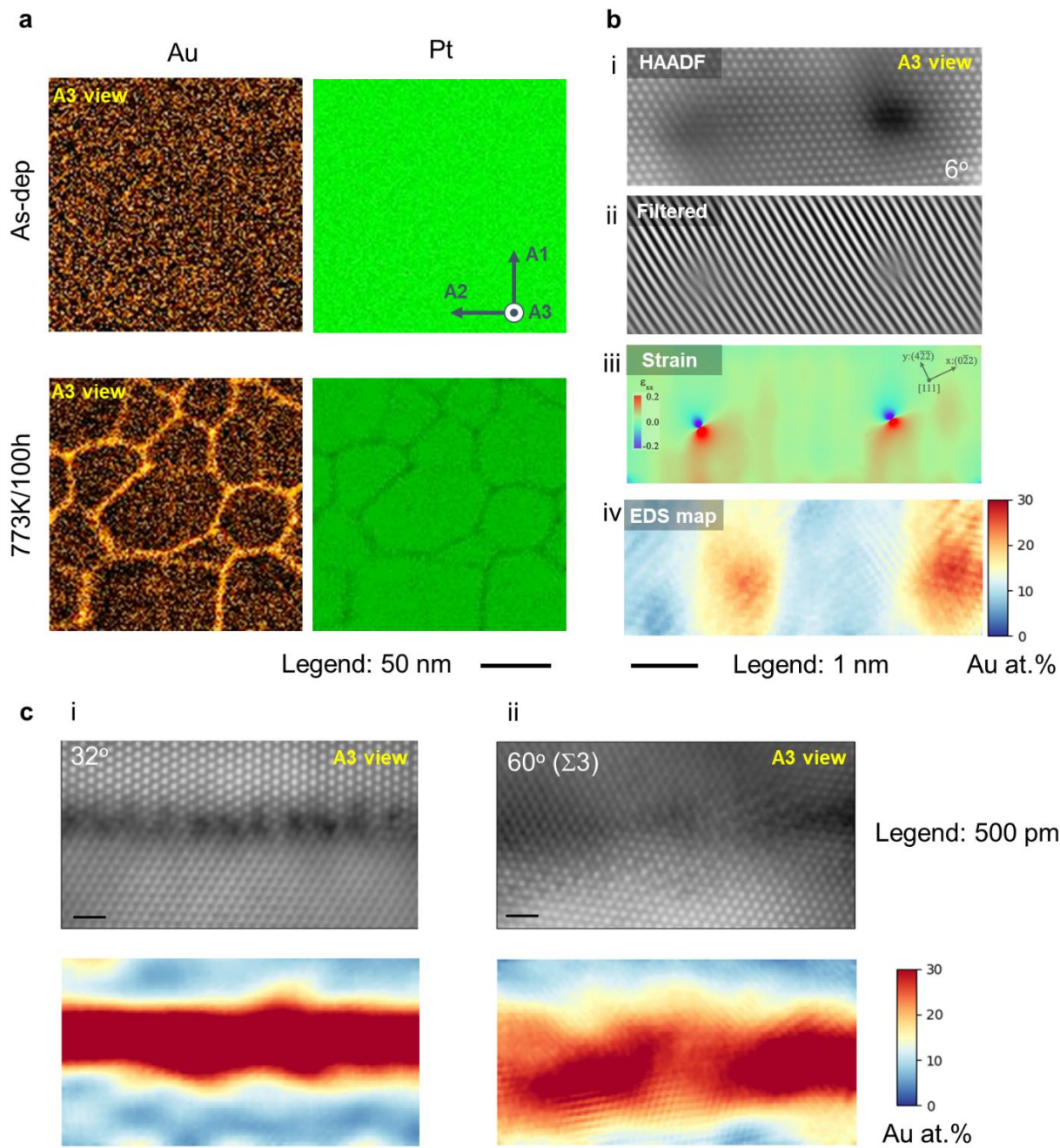

**Fig.1 Atomic scale correlation between structure and chemical composition a.** Nanoscale energy-dispersive X-ray spectroscopy (EDS) mapping of the as-deposited and 773K/100h annealed PtAu thin film. A1 and A2 are in-plane coordinates and A3 is an out-of-plane coordinate indicating the thin film growth direction. **b.** Scanning transmission electron microscopy (STEM) - EDS analyses of a low angle grain boundary (LAGB) with misorientation of 6° measured in the [111] zone axis. b i-iv show the high-angle annular dark- field (HAADF) image, Bragg-filtered phase map, $(0\bar{2}2)$ direction strain map, and the corresponding EDS map, respectively. **c.** HAADF images (top image in each sub-figure) and EDS maps (bottom image) for different regions of interest. c i-ii are from GBs with misorientation of 32° and 60° (Σ3), respectively.

**Grain boundary composition.** Fig. 2 elucidates the temporal evolution of the structure and composition of the PtAu film via cross-correlative TEM and APT studies. The as-deposited film appears to be chemically homogeneous, Fig. 2a-i. The atom probe crystallography [39] analysis, Fig. 2a-ii, indicates the existence of multiple grains within the collected APT volume. Cylindrical regions-of-interest positioned across the GBs quantify the local Au partitioning state. As shown in Fig. 2a-iii, the Au composition is approximately 5±3 at.% in both the grain interior and GB regions.

After 20h annealing at 773 K, Au partitions to defects. The bright-field image, Fig. 2b-i, and precession electron diffraction (PED) orientation map, Fig. 2b-ii, help identify the grain structure. Au segregates at both LAGBs and HAGBs, marked respectively in grey and black in Fig. 2b-ii. In the APT dataset, the GBs are revealed by a set of golden iso-composition surfaces encompassing regions segregated with over 8 at.% Au, Fig. 2b-iii&iv. Within the grain interior, we also note small clusters with a maximum Au content of 16 at.%. LAGBs contain linear patterns interpreted as chemically decorated dislocation arrays, as analyzed and simulated in detail in Ref. [32]. At HAGBs, inhomogeneous Au segregation appears across the interfaces, with random clusters distributed along the interface.

The bright-field image and the PED map of the 773K/100h PtAu thin film reveal dislocations, one LAGB and multiple HAGBs, Fig. 2c-i&ii. APT analysis, Fig. 2c-iii&iv, indicates that the grain interior only contains approximately 2 at.% Au, corresponding to the bulk solubility of Au in Pt at 773K [40], and the excess Au is hence segregated to microstructural defects. For example, individual dislocation lines, with an Au composition of approximately 20 at.%, were captured in the APT reconstructions, with the maximum content of Au reaching approximately 40 at.% in the 13° misorientation GB. Furthermore, we captured ripening and coarsening of Au clusters within the HAGBs of Fig. 2c [29].

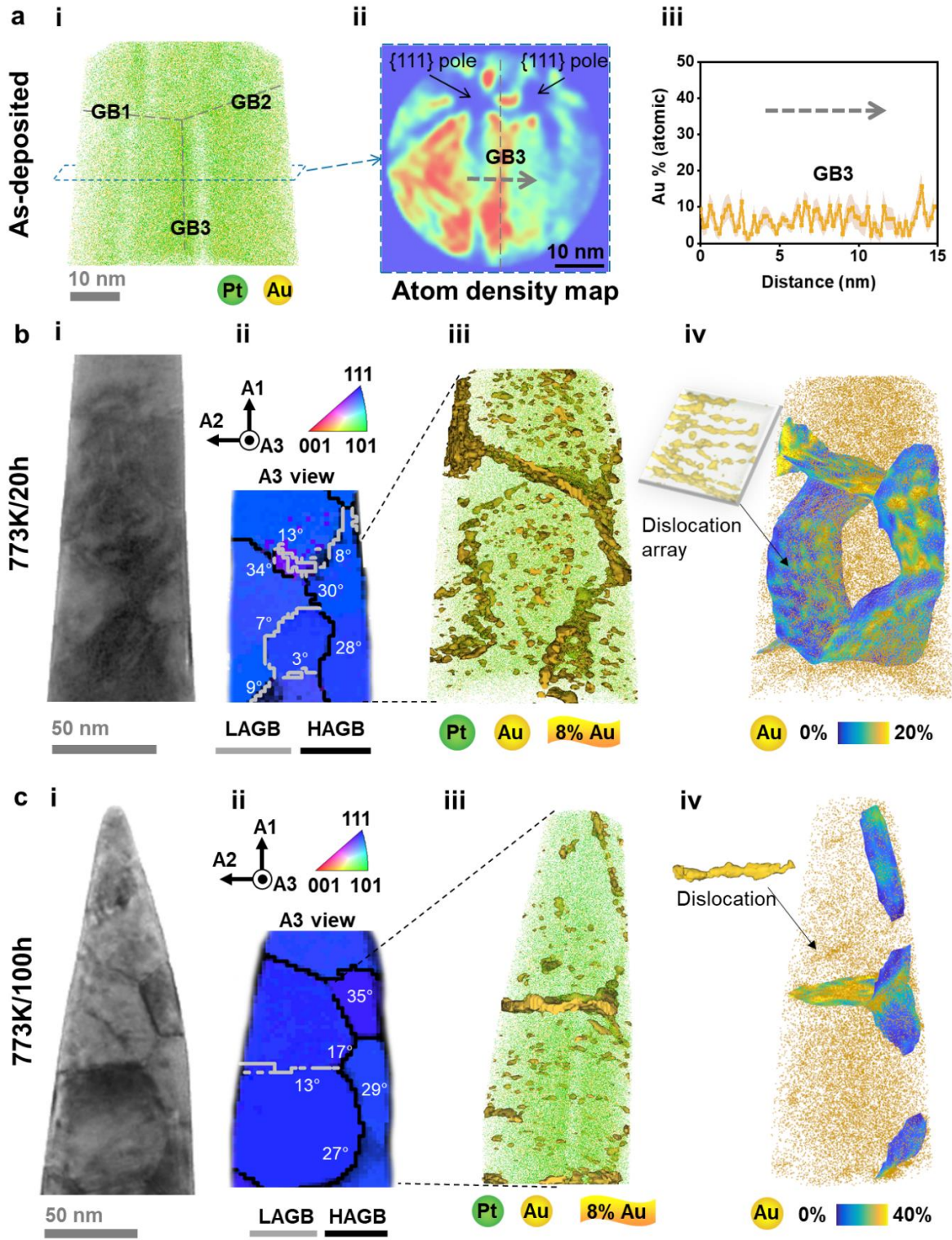

**Fig. 2 3D Correlative characterization of GB structure and chemical composition of the same material portion. a** Atom probe tomography (APT) analysis of the as-deposited PtAu thin film: **i** atom

map; **ii** 2D atom density map to show one GB; The {111} poles for adjacent grains are marked by black arrows. **iii** 1D composition profile across the GB shown in ii. Cross-correlative transmission electron microscopy (TEM) - precession electron diffraction (PED) and APT study of the **b** 773K/20h and **c** 773K/100h annealed PtAu thin films. The sub-figures for both b and c are **i** bright field image; **ii** orientation map; **iii** atom map with 8 at.% Au iso-composition surfaces embedded; **iv** atom map with overlaid GB in-plane chemical composition map. b-iv and c-iv also include the iso-composition surfaces to show an array of Au-decorated dislocations and one individual dislocation, respectively. Here, A1 and A2 are in-plane coordinates and A3 is an out-of-plane coordinate indicating the thin film growth direction.

Using our approach described in Ref. [41], we quantified the heterogeneous solute segregation at GBs. Extended Data Fig. 2 shows the histogram of the in-plane GB compositions measured in the APT data sets shown in Fig. 2. Here, the area fraction of the GB plane is plotted as a function of the Au content. From the histogram, we determine the highest composition of Au within the GBs directly. After up to 100 hours at 773K, the maximum Au content shifted to a higher value and the Au distribution changed from a Gaussian-like distribution in the as-deposited thin film to a broader distribution, indicating a spinodal-like segregation and decomposition behavior [29]. We previous reported low-dimensional ripening and cluster coarsening within GBs after extended annealing treatment [29], and discussed the concurrence and competition between segregation and phase separation under such topologically confined conditions.

**Electrochemical stability of Pt/PtAu thin films.** The electrochemical stability of PtAu alloy films have been probed by using a scanning flow cell coupled with inductively coupled plasma mass spectrometry (SFC-ICP-MS) [10]. As a benchmark material, we used a pure Pt thin film prepared in identical conditions. Considering that a fuel cell typically operates under almost steady-state conditions below 1.0 $V_{RHE}$ and the Pt cathode catalyst can locally be exposed to potentials up to 1.5 $V_{RHE}$ during startup or shutdown processes [42], the amount of dissolved Pt was evaluated during potential cycling with different upper potential limits (Fig. 3a). The cycling was performed at 10 mV/s from 0.1 $V_{RHE}$ to the upper potential limits (UPL) varied from 1.1 $V_{RHE}$ to 1.7 $V_{RHE}$ with the step size of 0.1 $V_{RHE}$. Each time two cycles were recorded. Such an electrochemical protocol has been previously suggested in the literature, and employed to probe dissolution of both Pt and Au electrodes [8-10,43,44]. Applying the same electrochemical procedure makes our results directly comparable to previous reports, and justifies the feasibility and transferability of our approach as a general method for catalyst stabilization.

Fig. 3b represents dissolution of a benchmark Pt thin film electrode. During the cycling of the potential within only the hydrogen adsorption and double layer regions, the concentration of Pt in the electrolyte remains below the detection limit of the ICP-MS. In line with previous reports [8-10], we did not detect dissolution of Pt at the region of potentials corresponding to adsorption of oxygen-containing species on Pt and their desorption. The dissolution of Pt becomes measurable in the cyclic voltamogramms with UPLs exceeding 1.2 $V_{RHE}$ (Fig. 3b). This observation corresponds to the literature data performed with higher resolution reporting onset of Pt dissolution at 1.15 $V_{RHE}$ [8-10]. With a further increase of the positive potential limit of the cyclic voltamogramms, the dissolution of Pt becomes more pronounced. Moreover, two peaks can be well distinguished in the dissolution profile for the cycles with the UPLs exceeding 1.5 $V_{RHE}$ [8-10]. The first and smaller peak corresponds to the dissolution of Pt during the anodic oxidation (in the positive going scan) and it is independent on the upper vertex potential. The second peak which dominates dissolution originates from the reduction of the surface in the decreasing section of the scan, starting below about +1.0 $V_{RHE}$. This peak, called cathodic dissolution, strongly depends on the nature of the anodically formed oxide film and, therefore, depends on the UPL.

When changing from the pure Pt electrode the three investigated PtAu alloys, the dissolution of both Pt and Au was monitored. Fig. 3c suggests that dissolution of Pt is decreasing in the as prepared, homogeneous PtAu alloy, which can be attributed to the reduced number of sites active towards dissolution and to reduced ability of the alloy to oxidation due to the presence of the more noble Au. Moreover, in contrast to the benchmark Pt electrode, here dissolution of Pt is below the detection limit up to the upper vertex potential of 1.3 $V_{RHE}$. The overall dissolution rate of Pt from the PtAu alloy is about 2 times lower compared to that measured for the benchmark Pt electrode (See Extended Data Table 1). Small peaks of Au start to appear on the dissolution profile at potentials above the onset for Au oxidation (above 1.3 $V_{RHE}$) [43,44] as revealed by the dissolution profile corresponding to the UPL of 1.4 $V_{RHE}$. The rate of Au dissolution grows with further increase of the positive potential. Interestingly, in contrast to Pt, it is impossible to distinguish between anodic and cathodic dissolution of Au even at higher vertex potentials, which can be attributed to the low content of Au in the alloy. Lower scan rates might be necessary to capture the difference between anodic and cathodic dissolution of Au, however, dissolution of Au itself is out of the scope of this work and will be addressed in a follow-up study.

After annealing treatments at 773K, the PtAu thin films exhibit a significant stabilization towards dissolution (Fig. 3d). PtAu alloy annealed during 20 h shows superior stability in transient conditions with dissolution of Pt and Au below the detection limit up to an upper potential limit of 1.5 $V_{RHE}$. These observed potential values are higher than the measured values of pure Pt or Au [45]. As the rate of Pt dissolution decreases the peaks for cathodic and anodic dissolution of Pt start to overlap, which indicates a decreasing number of active sites towards dissolution.

Further annealing to 100 hours leads only to a modest decrease in Pt dissolution rate at lower potentials compared to the as-prepared PtAu film, while the dissolution of Au becomes even higher at more positive potentials (Fig. 3e). For comparison, during two cycles with the upper potential limit of 1.7 $V_{RHE}$, the dissolution rate of Au for the 773K/100h sample is about 19.5 ng cm$^{-2}$ and exceeds the value of 9.1 ng cm$^{-2}$ observed for the 773K/20h sample. For comparison, Extended Data Table 1 lists the amounts of dissolved Au and Pt from each sample, which were calculated by integration of the dissolution profiles corresponding to two cycles at each upper potential limit.

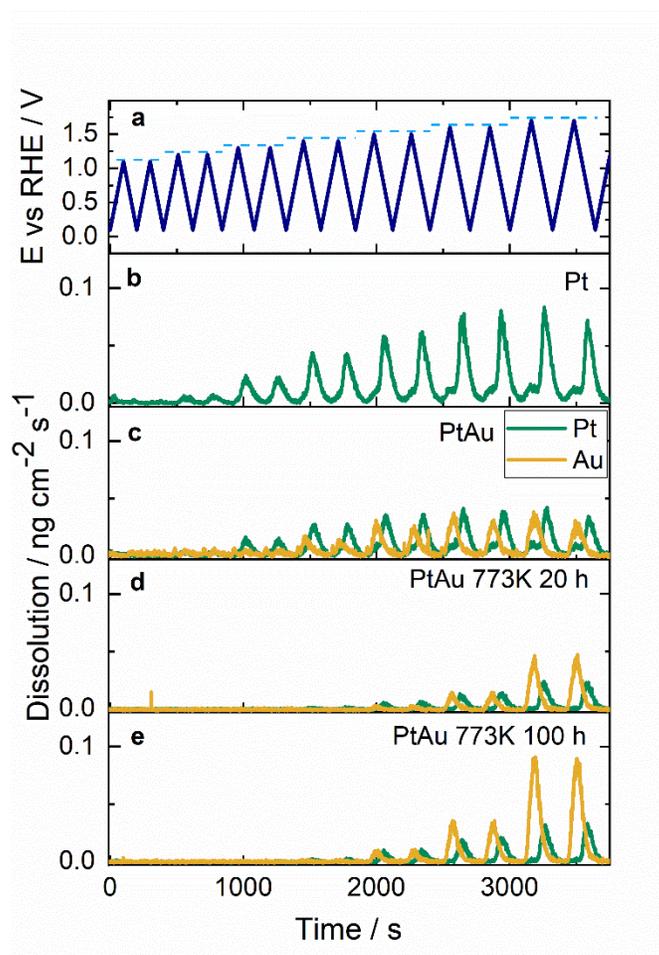

**Fig. 3 Stability of the Pt and PtAu thin film electrocatalysts. a** Applied electrochemical protocol that indicates cycling of electrode from 0.1 $V_{RHE}$ to the upper potential limits varying from 1.1 $V_{RHE}$ to 1.7 $V_{RHE}$ with the step size of 0.1 $V_{RHE}$. Scan rate 10 mV/s, electrolyte 0.1 M HClO$_4$. Each time two cycles were recorded. **b-e** Dissolution profiles of Pt and Au measured online upon application of above

mentioned electrochemical protocol to different electrodes: **b** benchmark Pt electrode, **c** as prepared PtAu, **d** PtAu 773K/20h, **e** PtAu 773K/100h. All potentials are presented in the reversible hydrogen electrode (RHE) scale. Slightly lower dissolution at the second cycle at UPL of 1.7 $V_{RHE}$ is attributed to the limitations of our measurement setup in conditions of intense oxygen evolution and formation of bubbles that may eventually block the surface of the working electrode.

**Discussion**

Fig. 4 schematically summarizes the relationship between stability and defect chemistry by comparing the dissolution process between the as-deposited and annealed PtAu thin films. In each figure, the transparent top layer in blue and the solid bottom layer in metallic color represent the chemical solution and the PtAu thin films, respectively. Local defects, such as GBs and dislocations, are highlighted in the thin film regions. According to the APT data the as-deposited PtAu film is a homogeneous alloy with uniform distribution of 5 at.% of Au in a matrix of Pt catalyst (Fig. 2a-i). Alloying of Pt with a low amount of Au leads to stability improvement by a factor of 2 and shift of the dissolution onset by 0.1 $V_{RHE}$ towards higher values. The shift of Pt transient anodic dissolution to more positive values in Pt-Au system has been previously reported. For example, Cherevko et al [46], observed stabilization of sub-monolayer Pt supported on Au after potential cycling and explained possible stabilization by alloying and shift of equilibrium due to lower concentration of the active sites. Recently, Lopes et al. reported reduced dissolution rates in Pt fuel cell catalyst provided by selective stabilization of low-coordinated Pt sites by Au [47].

In our case, during the stability test in transient conditions mimicking startup and shutdown of a fuel cell (1.0 $V_{RHE}$ – 1.5 $V_{RHE}$), the as-deposited PtAu film loses during each cycle a slightly higher amount of Pt than Au. Interestingly, at higher UPLs the accumulative amounts of dissolved Pt and Au equal to the dissolution of Pt from the benchmark electrode. This can be attributed to the oxygen evolution reaction (OER), which according to data reported by Kasian et al [15] and Danilovic et al [48], triggers enhanced dissolution of noble metals. The dissolution process of the as-deposited PtAu thin film is illustrated in Fig. 4a.

After annealing of PtAu, EDS (Fig. 1a) and APT (Fig. 2b&c) analyses clearly demonstrate Au segregation to a variety of microstructural defects, highlighted in golden in Fig. 4b. The PtAu alloy has been reported to be a nanocrystalline stabilized alloy [29,49]. We observe that segregation of Au solutes to defects not only helps to maintain the nanoscale grain size of the Pt host matrix after long-term annealing treatment (773K, up to 100 hours) (see Extended Data Fig. 1) but also improves the material's stability when exposed to catalytic reactions (Fig. 3). Solute segregation is a known effective strategy to stabilize nanocrystalline materials [50]. For example, Chookajorn *et al.* have demonstrated that the segregation of Ti at GBs substantially improved the nanocrystalline stability of a W-Ti alloy during high-temperature and long-term annealing [50]. Our PtAu alloy films show a similar trend also under harsh, catalytic exposure. The high level of Au segregation in the annealed PtAu alloy indeed explains the significant increase in nanostructure stability, which is evident from the dramatic increase in the potential at which dissolution can be detected in Fig. 3d.

Once Au atoms segregate into GBs, triggered by the annealing treatment during 20 h, the dissolution of Pt dramatically drops. This observation suggests that transient dissolution of Pt originates from lattice imperfections such as GBs and dislocations. GBs are generally locations of higher activity, they exhibit reduced coordination because not all bonds are saturated, causing an overall lower bond strength in such defect regions [1]. Enhanced dissolution from defects has been reported also for other noble metal catalysts, in particular for iridium [1] and its oxides [2]. Considering that Au is more stable and more noble than Pt, it becomes plausible that its segregation to the potentially unstable structural parts has beneficial impact on the catalyst's overall longevity, i.e. by passivating them. Thus, in the range of potentials corresponding to start-up and shut-down cycles in proton exchange membrane fuel cells (1.1 $V_{RHE}$ - 1.5 $V_{RHE}$) both elements Pt and Au exhibit only negligible dissolution. In one of the cycles, from 0.1 $V_{RHE}$ to 1.5 $V_{RHE}$, the loses of Pt and Au from a surface area of 1 $cm^2$ are equal to 0.45 ng and 0.2 ng, respectively. Considering the density of Pt atoms in the lattice and the atomic radii of Pt and Au it is possible to conclude that the observed dissolution of Pt corresponds to 0.15% of a single atomic layer. By comparison, dissolution of the benchmark Pt electrode exposed to an identical cycle corresponds to 2.2% of a monolayer, translating to a nearly 15 times improved longevity.

The overall Au content in the material is only 5 at.%, and APT clearly shows the segregation of Au atoms to defects, *i.e.*, GBs and dislocations, so that the local concentration Au in the GBs increases up to 20 at.% or more (Fig. 3b-iv). This means that the adsorption isotherm promotes the relaxation and trapping of the solute Au at those positions in the material, i.e. defects, where it is most needed to passivate them and hence protect the material against preferential dissolution. This means that even minute concentrations of dopants, through efficient partitioning between bulk and defect, can locally accumulate up to a substantial concentration that provides very effective protection. Under the oxygen evolution conditions, these Au atoms in the GBs will be subjected to the enhanced dissolution as schematically summarized in Fig. 4b.

Subsequent annealing treatment for a longer time, i.e. 100 hours, shows only minor stabilization. We attribute this finding to the phase separation, which leads to a dissolution behavior similar to that observed for the pure elements. Significantly higher dissolution rates of Au in such alloy will result in relatively fast complete removal of 5 at.% of the stabilizing element from the catalyst, leaving the Pt matrix unprotected, with a dissolution rate of similar or exceeding benchmark Pt electrode.

Taking a broader view, the correlated property measurements together with the atomic-scale characterization reveal that solute segregation at defects is of significant importance to the stability of catalysts. The quantification of such heterogeneous solute segregation at defects underpins the underlying understanding between local composition, defect structure and property relationships. The design strategy for superior catalysts should aim to segregate Au at defects particularly prone to act as preferential dissolution sites, thus playing a most effective role in mitigating the dissolution of Pt. The distribution of the decorating species, here the Au, to the different types of defects depends on the defect-specific trapping energies, values that are accessible through e.g. atomistic simulations. The heterogeneous redistribution of Au on the nanoscale to different defects provides an opportunity for the development of alloys, with dopants that are tailored to decorate defected engineered microstructures, for electrocatalytic applications.

We thereby opened here a new design avenue for the development of a new class of segregation-engineered catalytic materials. In future work, we will explore suited candidate doping elements with both, a high decoration and protection effect, across a wider compositional space, focusing particularly on less costly elements such as Ni and more stable alternatives such as Ti, W. The design will follow two main principles. First, we search for elements that have a strong preference for defect segregation. Second, these elements should then have a positive effect on the electrochemical stability without harming or even improving the material's activity.

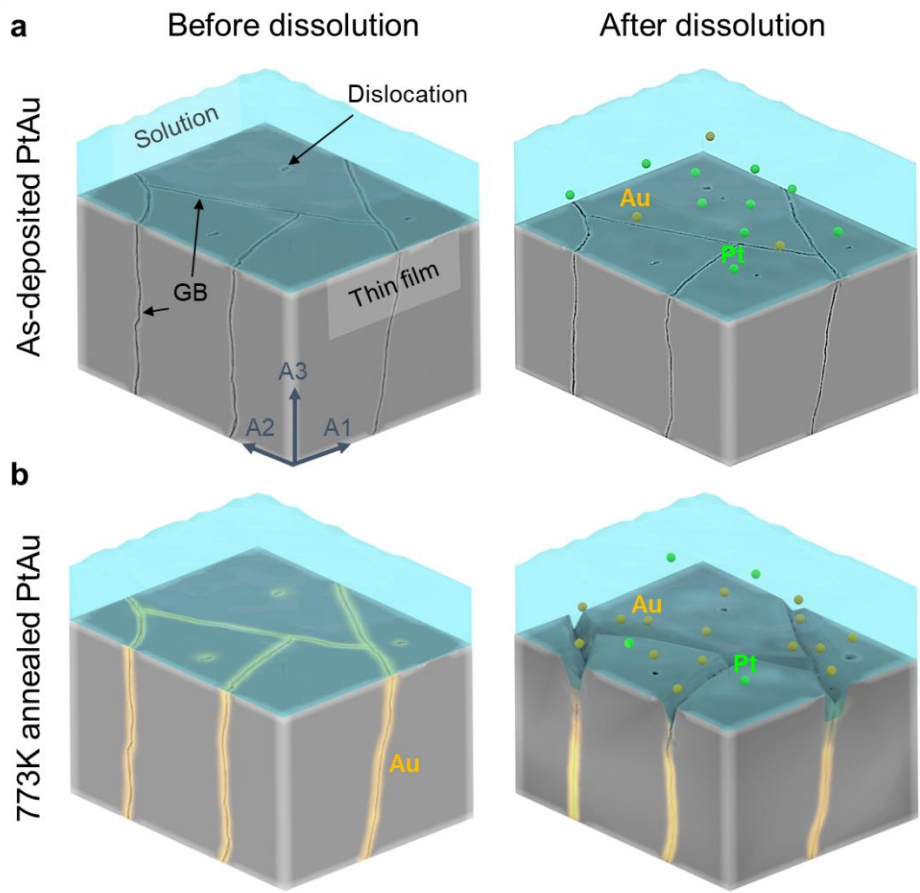

**Fig. 4 Schematic representation of the relationship between stability and defect chemistry. a** and **b** illustrate the dissolution processes of the as-deposited and annealed PtAu thin films, respectively. In each sub-image, the light blue region is the liquid solution and the metallic region is the PtAu thin film. The preferential dissolution of Au atoms occurs when Au atoms segregate to lattice defects, *i.e.*, GBs and dislocations, and act as stabilizing sites to prevent the dissolution of the Pt matrix atoms. A1 and A2 are in-plane coordinates and A3 is an out-of-plane coordinate indicating the thin film growth direction

**Conclusions**

We introduce segregation engineering as an approach to stabilize catalysts against dissolution via targeted chemical trapping of solutes at lattice defects. We provide detailed atomic-scale experimental findings on the relationship between the structural and chemical nature of lattice defects in a host catalyst and the decorated material's stability during electrocatalysis. Using annealing at 0.38 of the melting point for Pt, for 20 hours, we triggered the preferential segregation of over half of the 5 at.% of Au to a variety of microstructural defects, *i.e.*, GBs and dislocations, but avoided bulk phase separation. This tailored solute partitioning between bulk and defect regions improved the stability of Pt electrocatalysts against dissolution over a range of potentials in a typical fuel cell operation by more than an order of magnitude. This segregation engineering strategy opens new opportunities to the design of advanced catalysts for stable long-term energy conversion applications.

**Acknowledgments:** X.Z. and T.L. are supported by Alexander von Humboldt-Stiftung. X.Z. acknowledge funding by the German Research Foundation (DFG) for funding via project HE 7225/11-1. O.K. acknowledges funding from the German Research Foundation (DFG) – Project-ID 431791331 – SFB 1452. O.K. acknowledges support from the German Federal Ministry of Education and Research in the framework of the project CatLab (03EW0015A/B). G.B.T. gratefully acknowledges NSF- DMR-1709803 for support of this work. S.Z. acknowledges funding from the DFG within the framework of SPP 2370 (Project number 502202153). G.D. acknowledges support by the European Research Council under grant number 787446 GB-CORRELATE. The authors appreciated Dr. Thomas Koenig for sputtering thin film samples and the discussion with Dr. Christian Liebscher for electron microscopy analysis.

**Author contributions:** D.R., B.G., and O.K. conceived of the presented idea. X.Z. conducted the experiments and analytical characterization. O.K. performed the electrochemical stability data acquisition and analysis. B.G. guided atom probe analysis. T.K. and G.B.T provided the sample and technical interpretation of experimental findings. C.L., C.Z., S.L., and G.D. guided transmission electron microscopy analysis. S.Z. and S.H.K. assisted in the interpretation of the electrochemical results. X.Z., T.L., O.K., B.G., and D.R. wrote the paper. All authors provided critical feedback and helped shape the research, analysis, and manuscript.

**Competing interests:** Authors declare no competing interests.

**Data and materials availability:** All data is available in the main text or the Supplementary Information

**Methods**

**Sample preparation**

We homogenized Pt-Au alloy via co-sputtering Pt (99.9% pure, AJA, USA, at 100 W) and Au (99.9%, AJA, USA, at 5 W) targets onto 4-inch Si substrates with a 1.5 μm thermal $SiO_2$ diffusion and reaction barrier layer in an AJA 1300 stainless chamber. The base vacuum pressure prior to deposition was $<6\times10^{-6}$ Pa. During the sputtering the pressure was adjusted to 0.5 Pa. Before sputtering, a 10 nm Ti (at 50 W) was sputtered as an adhesion layer to increase the connection between the PtAu thin films and the 300 μm thick silicon [100] substrates. The film grew to an approximate thickness of 500 nm under the atmosphere of ultra-high purity argon that flowed as the working gas to maintain a pressure of 1.3 Pa. After sputtering, the substrates were cut into 1×2 $cm^2$ segments for annealing treatment and thermal stability quantification. The films were annealed at 773 K for 4, 20 or 100 hours in a custom vacuum chamber with a base pressure $< 1\times10^{-3}$ Pa. The temperature and duration of annealing treatment were selected to induce defect segregation and avoid phase separation.

The benchmark Pt thin film electrode was deposited by magnetron sputtering in Ar atmosphere (BESTEC GmbH, Berlin, Germany) at room temperature and 100 W. To prepare films with a minimal surface roughness, on the smooth substrates of single crystalline silicon [100] wafers with a 1.5 μm thermal $SiO_2$ diffusion and reaction barrier layer were used. The base vacuum before deposition was $2.5\times10^{-6}$ Pa. During the sputtering the pressure was adjusted to 0.5 Pa. The Ø3 inch target of Pt (99.9% pure, Evochem,
Germany) was pre-cleaned by sputtering prior to deposition. The resulting thickness of the obtained coating was approximately 200 nm.

**Electrochemical measurements**

We carried out all electrochemical measurements in 0.1M $HClO_4$ solution prepared by dilution of concentrated acid (Suprapur 70% HClO, Merck, Germany) in ultrapure water (PureLab Plus system, Elga, 18 MΩcm, total organic carbon < 3 ppb). The electrolyte was purged with Ar during the measurements. Online dissolution measurements were performed using a scanning flow cell connected to an ICP-MS (NexION 300X, Perkin Elmer) to investigate the stability of different Pt and PtAu thin films in real time. The set-up has been described in detail previously [1,51]. A Gamry potentiostat (Reference 600, USA) was used to control potential or current. A graphite rod served as counter electrode and saturated Ag/AgCl (Metrohm, Germany) was used as reference electrode. All indicated potentials refer to the potential of the RHE. The applied electrochemical protocol consisted from the measurement of the open circuit potential until the dissolution of Pt and Au from the surface oxides was stabilized. After that the potential was swept from 0.1 $V_{RHE}$ to the upper potential limits ranging from 1.1 $V_{RHE}$ to 1.7 $V_{RHE}$ with an increasing step of 0.1 $V_{RHE}$. Two cycles were recorded during each step. The stability was measured online by quantifying the amount of electrochemically dissolved Pt or Au during the applied electrochemical protocol by the ICPMS.

**TEM analysis**

The TEM samples were prepared in the plan-view orientation by focus ion beam (FIB) lift-out technique using FEI, Helios Nanolab 600i. During the process, the film wedge was rotated 90° to enable the TEM characterization to be performed along the direction of the columnar grains to minimize the grain overlap [29,32]. The grain orientations were measured by PED [52] using the commercial package NanoMEGAS ASTAR platform that runs in a STEM, JEOL JEM-2200FS that operates at 200keV. The PED scans operated with a 0.3° precession angle with 0.5 nm spot size and 3 nm step size over 2100×600 nm$^2$ regions of interest (ROI) [53]. The scanning data was converted for grain analysis using the TSL OIM Analysis 8 software package.

Nanoscale EDS was also performed in the same JEOL JEM-2200FS microscope. We collected HR-STEM and EDS data in a Cs probe-corrected FEI Titan Themis 60-300 operated at 300 kV with a semi-convergence angle of 23.6 mrad and a semi-collection angle of 103 to 220 mrad. The probe current and the probe spacing were set to be 100 pA and 24.9 pm, respectively. The HR-STEM & EDS data was acquired in a series of at least 400 frames each recorded with a dwell time of 20 μs. The collected images were stacked and corrected with the help of non-rigid registration (NRR) to minimise scan distortion and non-local principal component analysis (NLPCA) to increase signal-to-noise ratio [54-58]. The details of the entire data process are documented in the Jupiter notebook – PyEDS - and published in GitHub (Link: https://github.com/RhettZhou/pyEDS).

**APT analysis**

We characterize the composition and element distribution of the alloy films by APT in a Cameca Instruments Local Electrode Atom Probe (LEAP) 5000XS. The required needle-shaped geometry for APT specimens was prepared using a FIB lift-out and annular milling technique [28,59,60] by using FEI, Helios Nanolab 600i. Similar to the plan-view TEM sample preparation, the lift-out wedge was rotated 90° to insure the TEM characterization along the direction of the columnar grains [60]. The FIB extracted tips were mounted onto a half Mo grid which has been prior sharpened by an FEI Helios plasma focused ion beam (PFIB) using of a Xe source. Before the field evaporation, the APT specimen was imaged in a JEOL JEM-2200FS TEM for PED grain-to-grain mapping. The beam was also precessed at 0.3° at a scanning step size of 2.5 nm. At last, the APT specimens were field evaporated in the LEAP 5000 XS that operated at a specimen set point temperature of 40 K with a laser pulse energy between 250 to 350 pJ and a pulse repetition rate of 200 kHz for a 1.2% atoms per pulse detection rate. AP Suite 6.1 software platform was employed to reconstruct the collected ions [27,61]. We employed GB composition mapping method to visualize the quantify the solute segregation within the GB planes [30,31].

**Extended Data Fig. 1 | Microstructure characterization. a** Grain orientation and grain boundary (GB) maps for the as-deposited **i** Pt and **ii** PtAu, and **iii** the 773 K/100 h PtAu thin films. A1 and A2 are in-plane coordinates and A3 is an out-of-plane coordinate indicating the thin film growth direction. **b** cumulative area fraction of the grain size. **c** histogram plots of the GB types. **d** Pole figures from the **i** Pt and **ii** PtAu, and **iii** the 773K/100h PtAu thin films.

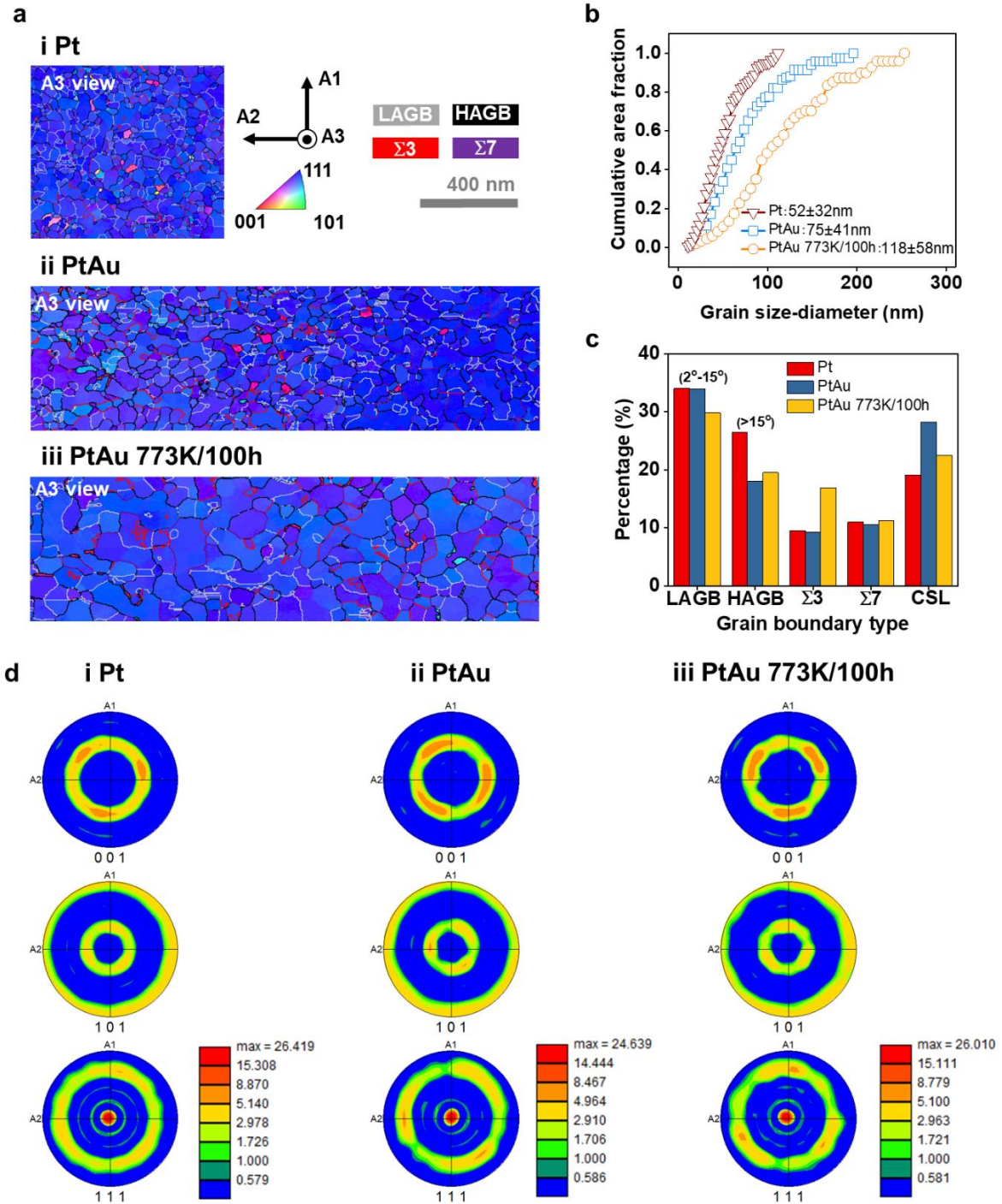

**Extended Data Fig. 2 | Quantification of the chemical composition of grain boundaries.** The histograms of the in-plane grain boundary (GB) compositions measured from atom probe tomography data sets in Fig. 2, showing the area fraction of the GB plane as a function of the Au content.

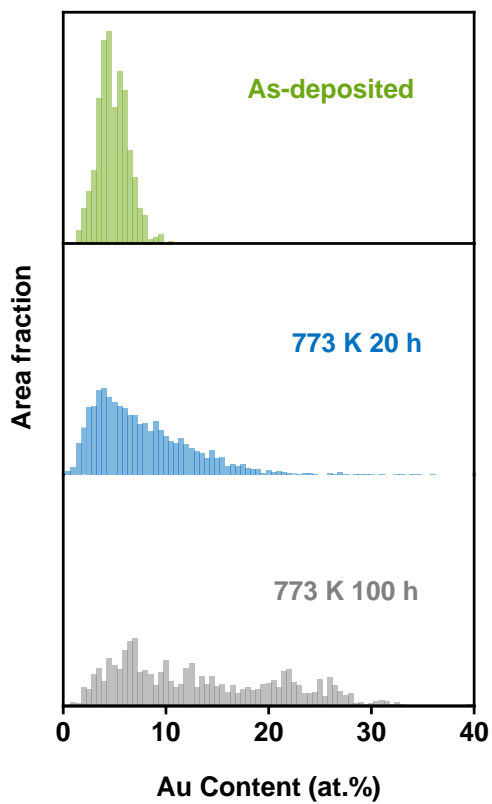

**Extended Data Table 1 | Electrochemical dissolution of Pt and Au from different samples**

| Electrode | Pt | PtAu | | PtAu 773K 20 h | | PtAu 773K 100 h | |
|---|---|---|---|---|---|---|---|
| | Dissolved in two cycles*, ng cm$^{-2}$ | | | | | | |
| Applied UPL, $V_{RHE}$ | Pt | Pt | Au | Pt | Au | Pt | Au |
| 1.1 | - | - | - | - | - | - | - |
| 1.2 | 1.1 | - | - | - | - | - | - |
| 1.3 | 4.5 | 2.8 | - | - | - | - | - |
| 1.4 | 8.7 | 5.5 | 3.4 | - | - | - | - |
| 1.5 | 12.5 | 7.4 | 6.1 | 0.9 | 0.4 | 1.6 | 1.6 |
| 1.6 | 15.6 | 8.5 | 7.7 | 2.3 | 2.4 | 3.7 | 6.6 |
| 1.7 | 16.0 | 8.4 | 7.6 | 4.0 | 9.1 | 5.9 | 19.5 |
| Totally dissolved during the protocol | 58.4 | 32.6 | 24.8 | 7.2 | 11.9 | 11.2 | 27.7 |

*The amount of dissolved Pt and Au was calculated by integration of the dissolution profiles in Fig. 3.